\documentclass[aps,showpacs,tightenlines,twocolumn,nofootinbib,nobibnotes,superscriptaddress]{revtex4-1}
\usepackage{amsmath,amssymb,amsfonts,bm}
\usepackage{graphicx}
\usepackage{epstopdf}
\usepackage{dcolumn}
\usepackage{mathrsfs}
\usepackage{tgtermes}
\usepackage[colorlinks=true,linkcolor=blue,citecolor=blue, urlcolor=blue,bookmarks=false]{hyperref}
\usepackage{lipsum}
\bibpunct{[}{]}{,}{n}{}{}
\usepackage{multirow}
\begin{document}
\title{Density-driven higher-order topological phase transitions in amorphous solids}
\date{\today }
\author{Tan Peng}

\affiliation{School of Mathematics, Physics and Optoelectronic Engineering, Hubei University of Automotive Technology, Shiyan 442002, China}

\affiliation{Department of Physics, Hubei University, Wuhan 430062, China}

\author{Chun-Bo Hua}
\affiliation{School of Electronic and Information Engineering, Hubei University of Science and Technology, Xianning 437100, China}

\affiliation{Department of Physics, Hubei University, Wuhan 430062, China}

\author{Rui Chen}
\affiliation{Department of Physics, The University of Hong Kong, Pokfulam Road, Hong Kong 999077, China}

\author{Zheng-Rong Liu}
\affiliation{Department of Physics, Hubei University, Wuhan 430062, China}

\author{Hai-Ming Huang}
\affiliation{School of Mathematics, Physics and Optoelectronic Engineering, Hubei University of Automotive Technology, Shiyan 442002, China}

\author{Bin Zhou}\email{binzhou@hubu.edu.cn}
\affiliation{Department of Physics, Hubei University, Wuhan 430062, China}

\begin{abstract}
Amorphous topological states, which are independent of the specific spatial distribution of microscopic constructions, have gained much attention.
Recently, higher-order topological insulators, which are a new class of topological phases of matter, have been proposed in amorphous systems. Here, we propose a density-driven higher-order topological phase transition in a two-dimensional amorphous system. We demonstrate that the amorphous system hosts a topological trivial phase at low density. With an increase in the density of lattice sites, the topological trivial phase converts to a higher-order topological phase characterized by a quantized quadrupole moment and the existence of topological corner states. Furthermore, we confirm that the density-driven higher-order topological phase transition is size dependent. In addition, our results should be general and equally applicable to three-dimensional amorphous systems. Our findings may greatly enrich the study of higher-order topological states in amorphous systems.

\end{abstract}

\maketitle

\section{Introduction}
Amorphous solids, which can be grown by most solids, are ubiquitous in condensed matter \cite{zallen2008physics}. Unlike traditional crystalline solids, amorphous solids lack long-range order but maintain short-range order due to the random arrangement of internal atoms. Recently, amorphous topological states, such as topological insulators \cite{PhysRevLett.118.236402,PhysRevB.96.121405,mitchell2018amorphous,bourne2018non,
PhysRevB.99.045307,PhysRevB.99.165413,Chern_2019,costa2019toward,PhysRevB.101.035142,PhysRevResearch.2.013053,
PhysRevResearch.2.043301,Marsal30260,peiheng2020photonic,grushin2021topological,Focassio_2021,PhysRevB.103.214203}, topological superconductors \cite{poyhonen2018amorphous}, and topological metals \cite{PhysRevLett.123.076401}, have been theoretically proposed in amorphous systems. Meanwhile, several experimental works have reported observations of topological states in amorphous materials, including silica bilayers \cite{Buchner,Burson_2016} and Bi$_{2}$Se$_{3}$ thin films \cite{corbae2021evidence}. Amorphous topological states, which are independent of the specific spatial distribution of the microscopic constructions, have become a hot research topic in the pursuit of topological states of matter in aperiodic systems with a rapidly growing number of novel proposals.

Recently, higher-order topological insulators, which are a new class of topological phases of matter, have been proposed in various systems \cite{PhysRevB.97.205135,
Fangeaat2374,xu2017topological,PhysRevLett.120.026801,PhysRevLett.121.116801,PhysRevB.97.155305,PhysRevB.97.205136,PhysRevB.97.241402,
PhysRevB.97.241405,PhysRevB.98.045125,PhysRevLett.121.096803,PhysRevLett.121.186801,PhysRevB.97.094508,PhysRevB.98.235102,
PhysRevB.98.241103,PhysRevB.98.245102,PhysRevLett.123.016805,PhysRevLett.122.076801,PhysRevLett.122.204301,PhysRevLett.123.247401,
PhysRevLett.123.256402,PhysRevB.99.245151,PhysRevB.100.085138,PhysRevB.100.235302,PhysRevB.101.041404,PhysRevLett.122.236401,
PhysRevB.99.125149,PhysRevB.100.205406,PhysRevLett.125.097001,
PhysRevB.98.081110,PhysRevB.98.165144,PhysRevB.98.201114,PhysRevLett.124.216601,PhysRevX.9.011012,PhysRevLett.122.086804,
PhysRevLett.123.156801,PhysRevLett.123.167001,PhysRevLett.123.177001,PhysRevLett.123.186401,PhysRevLett.123.216803,PhysRevB.99.041301,
PhysRevB.99.235132,PhysRevResearch.3.013239,PhysRevB.102.094503,wieder2020strong,PhysRevResearch.2.033029,schindler2018higher,
serra2018observation,xue2019acoustic,ni2019observation,peterson2018quantized,mittal2019photonic,zhang2020low,noh2018topological,imhof2018topolectrical,
PhysRevB.100.201406,PhysRevLett.123.196401,PhysRevLett.124.036803,PhysRevB.102.241102,PhysRevResearch.2.033071,lv2021realization}. Higher-order topological insulators, that are established in crystalline systems with crystalline symmetries, have lower dimensional boundary states compared with conventional topological insulators \cite{saha2021higher,schindler2021tutorial,xie2021higher,Benalcazar61,PhysRevB.96.245115,Schindlereaat0346,PhysRevLett.119.246401,PhysRevLett.119.246402,PhysRevB.92.085126}. Surprisingly, despite the lack of spatial symmetry, amorphous systems can still host higher-order topological phases \cite{PhysRevResearch.2.012067,PhysRevLett.126.206404}. For instance, Wang \emph{et al}. have demonstrated that a second-order topological insulating phase can exist in a three-dimensional amorphous system without any spatial order \cite{PhysRevLett.126.206404}.

\begin{figure}[tp]
	\includegraphics[width=8.5cm]{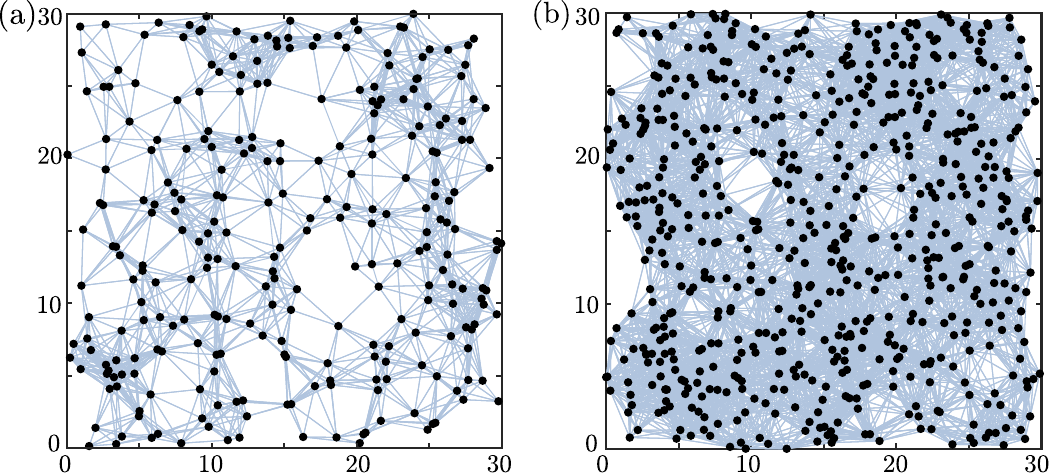} \caption{Schematic diagram of a random lattice with (a) $270$ sites ($\rho=0.3$) and (b) $630$ sites ($\rho=0.7$) marked by black points. The sample size is $30\times30$. }%
\label{fig1}
\end{figure}

It is worth noting that some previous works have demonstrated that amorphous topological phase transitions are related to density \cite{PhysRevLett.118.236402, poyhonen2018amorphous,PhysRevResearch.2.013053} . Sahlberg \emph{et al}.
proposed a scaling theory of an amorphous topological phase transition from a topological trivial phase to a topological Chern insulator phase driven by the density of lattice sites \cite{PhysRevResearch.2.013053}. In addition, McMillan \emph{et al}. presented experimental evidence for the occurrence of a density-driven phase transition between semiconducting and metallic polyamorphs of silicon by changing pressure \cite{mcmillan2005density}. However, whether a topological phase transitions from a topological trivial phase to a higher-order topological phase driven by density can exist in amorphous systems is unclear.

In this paper, we investigate the density-driven higher-order topological phase transition in a two-dimensional ($2$D) amorphous system without any spatial order. The zero-energy corner modes are protected by particle-hole symmetry and effective chiral symmetry which is similar to Ref. \cite{PhysRevResearch.2.012067}. By calculating the quadrupole moment and the probability density of the in-gap eigenstates, it is found that the amorphous system hosts a topological trivial phase when the density is below the critical density, beyond which the system holds a higher-order topological phase characterized by a quantized quadrupole moment and the existence of corner states. Our results should be general, and also applicable to three-dimensional ($3$D) amorphous systems.

The rest of the paper is organized as follows. In Sec.~\ref{Models}, we introduce the higher-order topological insulator in a $2$D random lattice and give the details of our numerical methods. Then, we provide the numerical results for studying a higher-order topological phase transition driven by density in $2$D and $3$D random lattices in Sec.~\ref{NS}. Finally, we summarize our conclusions in Sec.~\ref{CD}.

\section{Models and Method}
\label{Models}

\begin{figure}[tp]
    \centering
	\includegraphics[width=8.5cm]{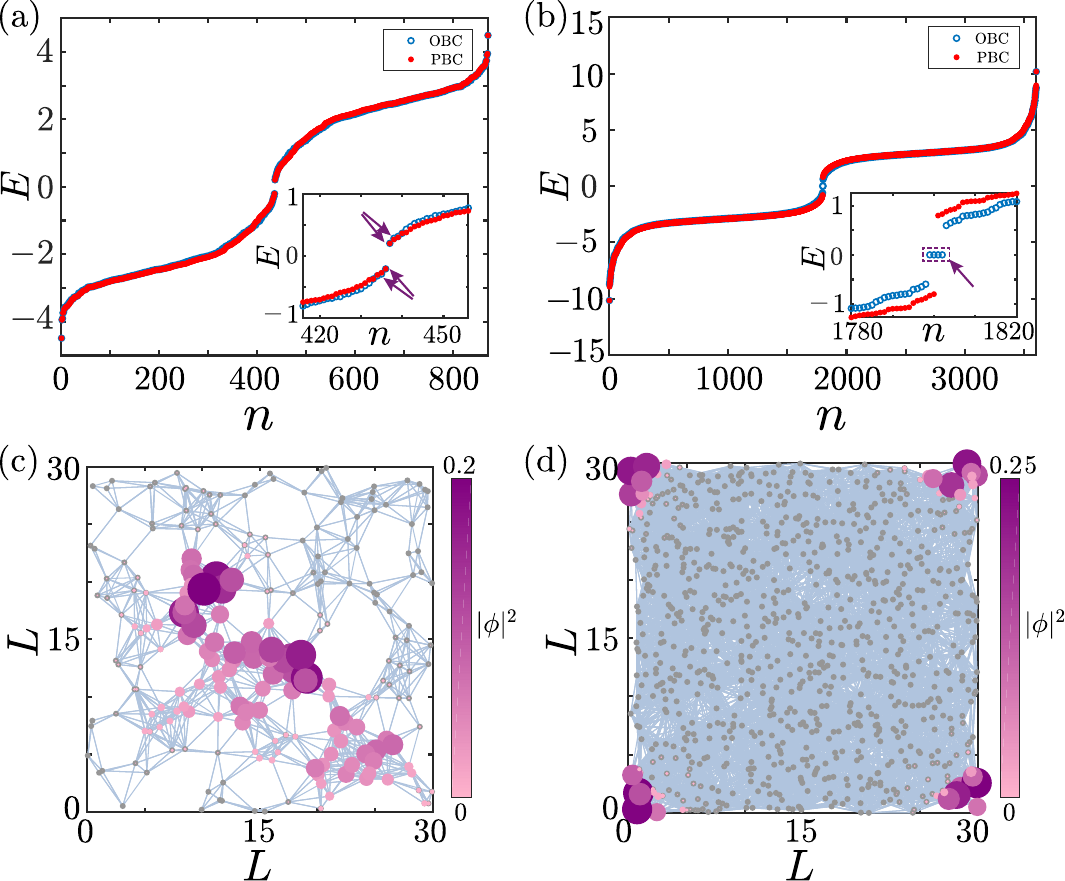} \caption{Energy spectrum (top) and the probability density (bottom) of the four eigenstates which are the nearest to zero energy (marked by purple arrows) at an open boundary condition (OBC) with different densities. (a), (c) $\rho=0.24$ and (b), (d) $\rho=1$. The Dirac mass is set as $M=0$. The sample size is $30\times30$.}%
\label{fig2}
\end{figure}

First, we construct a random lattice by placing $N$ sites randomly in an $L \times L$ region, where the positions of the sites are sampled from an uncorrelated uniform distribution, as shown in  Fig.~\ref{fig1}. We define a cutoff distance $R=4$ such that hoppings (marked by light blue lines in
Fig.~\ref{fig1} ) for $|r_{jk}| < R$ exist for each site, where $r_{jk}$ is the distance from site $j$ to site $k$. It is noted that the density of the lattice sites $\rho=N/V$, where $V=L\times L$ is the volume of the sample. Next, we realize a higher-order topological insulator in an amorphous system by adding a proper mass term into the quantum spin Hall (QSH) insulator to gap the topological edge states, leading to the appearance of topological corner states \cite{PhysRevLett.124.036803}. The model Hamiltonian is given by
\begin{equation}
H=H_{0}+H_{m},
\label{H}
\end{equation}
with the QSH insulator Hamiltonian
\begin{eqnarray}
H_{0} &=&-\sum_{j\neq k}\frac{l(r_{jk})}{2}c_{j}^{\dagger }[it_{1}(s_{3}\tau
_{1}\cos \psi _{jk}+s_{0}\tau _{2}\sin \psi _{jk}) \notag \\
&&+t_{2}s_{0}\tau _{3}]c_{k}+\sum_{j}(M+2t_{2})c_{j}^{\dagger }s_{0}\tau _{3}c_{j},
\label{H0}
\end{eqnarray}
and the mass term
\begin{equation}
H_{m}=g \sum_{j\neq k}\frac{l(r_{jk})}{2}c_{j}^{\dagger } s_{1}\tau _{1}\cos (\xi \psi _{jk})c_{k},
\label{Hm}
\end{equation}
where  $c_{j}^{\dag }=(c_{j\alpha \uparrow }^{\dag },c_{j\alpha \downarrow }^{\dag
},c_{j\beta \uparrow }^{\dag },c_{j\beta \downarrow }^{\dag })$  is the creation  operator at site $j$. Each site contains four degrees of freedom. $\alpha$ and $\beta$ present two orbitals at each site. $\uparrow$ and $\downarrow$ denote spin up and spin down, respectively. $j$ and $k$ denote lattice sites running from $1$ to $N$, and $N$ is the total number of lattice sites. $s_{1,2,3}$ and $\tau_{1,2,3}$ are the Pauli matrices acting on the spin and orbital spaces, respectively. $s_{0}$ and $\tau_{0}$ are the $2\times 2$ identity matrices. $\psi_{jk}$ is the polar angle of the bond between site $j$ and $k$ with respect to the horizontal direction. $l(r_{jk})=e^{1-r_{jk}/\lambda }$ is the spatial decay factor of hopping amplitudes with the decay length $\lambda $. $M$ is the Dirac mass. $t_{1}$ and $t_{2}$ are the hopping amplitudes. Without loss of generality, we set $t_{1}=1$, $t_{2}=1$, and $\lambda =1$ for simplicity. $\xi $ is the varying period of the mass term. For a square sample, we set $\xi = 2$. The Wilson mass term $H_{m}$, relaying in the polar angle of the bond $\psi_{jk}$, can result in an effective edge mass domain structure. The corner states appear only when a mass domain wall forms, which can be explained by a generalized Jackiw-Rebbi mechanism \cite{PhysRevD.13.3398}. The Hamiltonian $H$ in Eq.(\ref{H}) respects particle-hole symmetry $P=s_{3}\tau_{1}K$ and effective chiral symmetry $S_{\text{eff}}=s_{2}\tau_{1}$, where $K$ is the complex conjugate operator.

The random lattice lacks translation invariance, thus, to characterize the higher-order topological phases of the random lattice, we will employ the real-space topological invariant quadrupole moment, which has been proposed in two previous works \cite{PhysRevB.100.245134,PhysRevB.100.245135}. The real-space quadrupole moment is given by \cite{PhysRevB.100.245134,PhysRevB.100.245135,PhysRevB.101.195309,PhysRevLett.125.166801,PhysRevResearch.2.012067}
\begin{equation}
q_{xy}=\frac{1}{2\pi }{\rm{Im}} \log [\det (\Psi _{\rm{occ}}^{\dagger }\hat{U}\Psi
_{\rm{occ}})\sqrt{\det (\hat{U}^{\dagger })}],
 \label{qxy}
\end{equation}
where $\hat{U}\equiv \exp [i2\pi \hat{X}\hat{Y}/N]$ with $\hat{X}$ ($\hat{Y}$) the position operator. The matrix $\Psi _{\rm{occ}}$ is the eigenvectors of occupied states, such that $\Psi _{\rm{occ}}\Psi_{\rm{occ}}^{\dagger }$ is the projector to the occupied subspace. The case with $q_{xy}=0$ corresponds to the topological trivial phase, and $q_{xy}=0.5$ corresponds to the higher-order topological phase. It is noted that Ono \emph{et al}. indicated that the validity of the real-space quadrupole moment proposed by two previous works is still under discussion \cite{PhysRevB.100.245133}. In fact, to define a satisfactory real-space quadrupole moment is a difficult task. Despite the fact that the real-space formula of the bulk quadrupole moment has been applied to the study of higher-order topological insulators, it is worth defining a satisfactory formulation of the bulk quadrupole moment in future works. In addition, another appropriate way to characterize the higher-order topological phase is to adopt the existence of corner states as a working definition \cite{PhysRevB.99.085406,PhysRevLett.126.146802,PhysRevB.104.245302}. In the following calculation, we employ the quantized quadrupole moment ($q_{xy}=0.5$) as well as the existence of corner states as criteria to characterize the higher-order topological phase. Note that it has been proved that the quantization of the quadrupole moment can be protected by chiral symmetry and particle-hole symmetry \cite{PhysRevLett.125.166801}.

\begin{figure}[tp]
\centering
	\includegraphics[width=8.5cm]{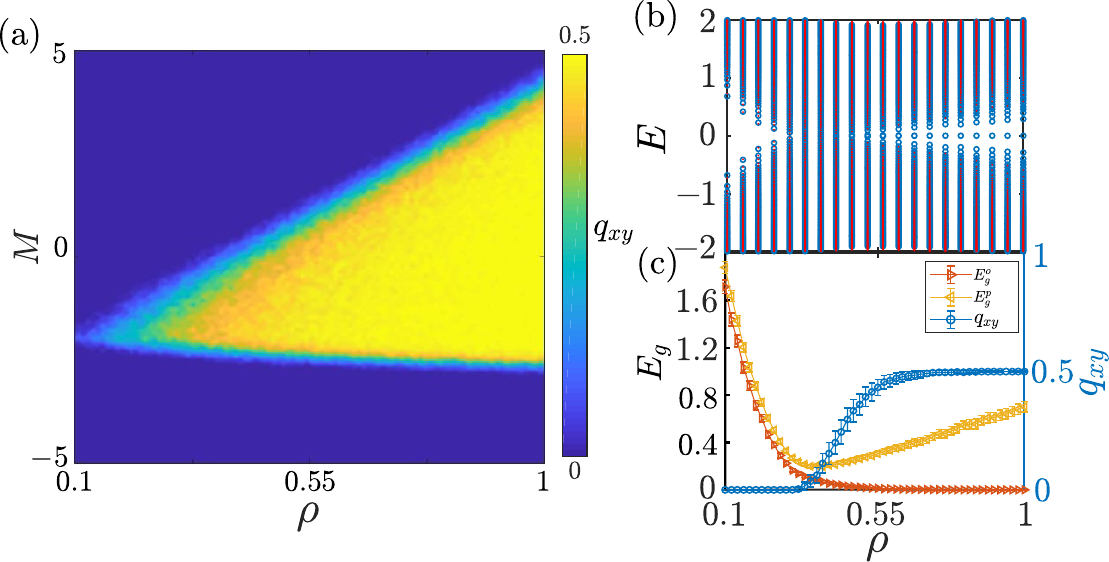}\caption{(a) Topological phase diagram of the amorphous lattice in ($\rho, M$) space obtained by calculating the topological invariant quadrupole moment $q_{xy}$ with $g=1$.  Periodic boundary conditions are taken in our calculation. (b) Energy spectrum vs density $\rho$ with $M=0$ in the OBC  marked by the blue circles and the PBC marked by the red dots. (c) Energy gap and quadrupole moment $q_{xy}$ vs density $\rho$ with $M=0$. The system with OBC marked by the red line shows that the edge energy gap becomes smaller with an increase of $\rho$ and eventually closes. The system with PBC marked by the yellow line shows that the bulk energy gap has undergone the process of closing and reopening. The sample size is $30\times30$ for (a)-(c). An average of some $100$ random configurations are performed for (a) and (b), and an average of $1000$  random configurations in (c).
} %
\label{fig3}
\end{figure}

\section{Numerical simulation}
\label{NS}

We map out the energy spectrum of the amorphous lattice with different densities of both the open boundary condition (OBC) (marked by blue circles) and periodic boundary condition (PBC) (marked by red dots) in Figs.~\ref{fig2}(a) and \ref{fig2}(b), where $g = 1$. When the density is small [$\rho=0.24$ in Fig.~\ref{fig2}(a)], the system is a topological trivial phase accompanied by the appearance of an energy gap in both OBC and PBC. The corresponding probability density of the middle four eigenstates which are the nearest to the zero-energy [marked by purple arrows in Fig.~\ref{fig2}(a)] are localized in the bulk [see Fig.~\ref{fig2}(c)]. When the density $\rho=1$, the system hosts a higher-order topological phase with four degenerate zero energy states localized at the four corners of the lattice [see Fig.~\ref{fig2}(d)], indicating that the system undergoes a phase transition from a normal insulator phase to a higher-order topological phase with an increase in density. It is noted that the existence of the corner states is strong evidence for the emergence of the higher-order topological phase.

To further study the effect of density on higher-order topological phase transitions, we plot the real-space quadrupole moment as a function of density $\rho$ and the Dirac mass $M$ in Fig.~\ref{fig3}(a). The color map represents the magnitude of the real-space quadrupole moment $q_{xy}$. It is found that the system is in a topological trivial phase characterized by $q_{xy}=0$ when the density is small ($\rho\approx0.1$) with arbitrary Dirac mass $M$. With an increase in density, the trivial phase converts to a higher-order topological phase with $q_{xy}$ changing from $0$ to $0.5$. However, the critical density points where the phase transitions occur are dependent on Dirac mass $M$. Figure~\ref{fig3}(b) shows the energy spectrum versus density $\rho$ at $M=0$ with OBC (marked by the blue circles) and PBC (marked by the red dots), respectively. It is noted that the energy gap calculated in the PBC corresponds to the bulk energy gap. It is found that the system is a normal insulator with a large bulk energy gap when the density is small. With an increase in density, the bulk energy gap is closed at $\rho \approx 0.28$. However, when the density continues to increase to $\rho \approx 0.4$, the bulk energy gap is reopened. Simultaneously, a series of zero-energy modes appear in the energy gap. It is demonstrated that the system undergoes a topological phase transition from a normal insulator to a higher-order topological insulator. In Fig.~\ref{fig3}(c), we plot the quadrupole moment and the energy gap versus density $\rho$ with $M=0$. It is found that the system hosts a trivial phase with $q_{xy}=0$ when the density is small ($0.1<\rho<0.28$). With an increase in density, $q_{xy}$ increases and is finally quantized to $0.5$ ($\rho\approx 0.7$), indicating that the system hosts a higher-order topological phase.

\begin{figure}[tp]
	\includegraphics[width=8.5cm]{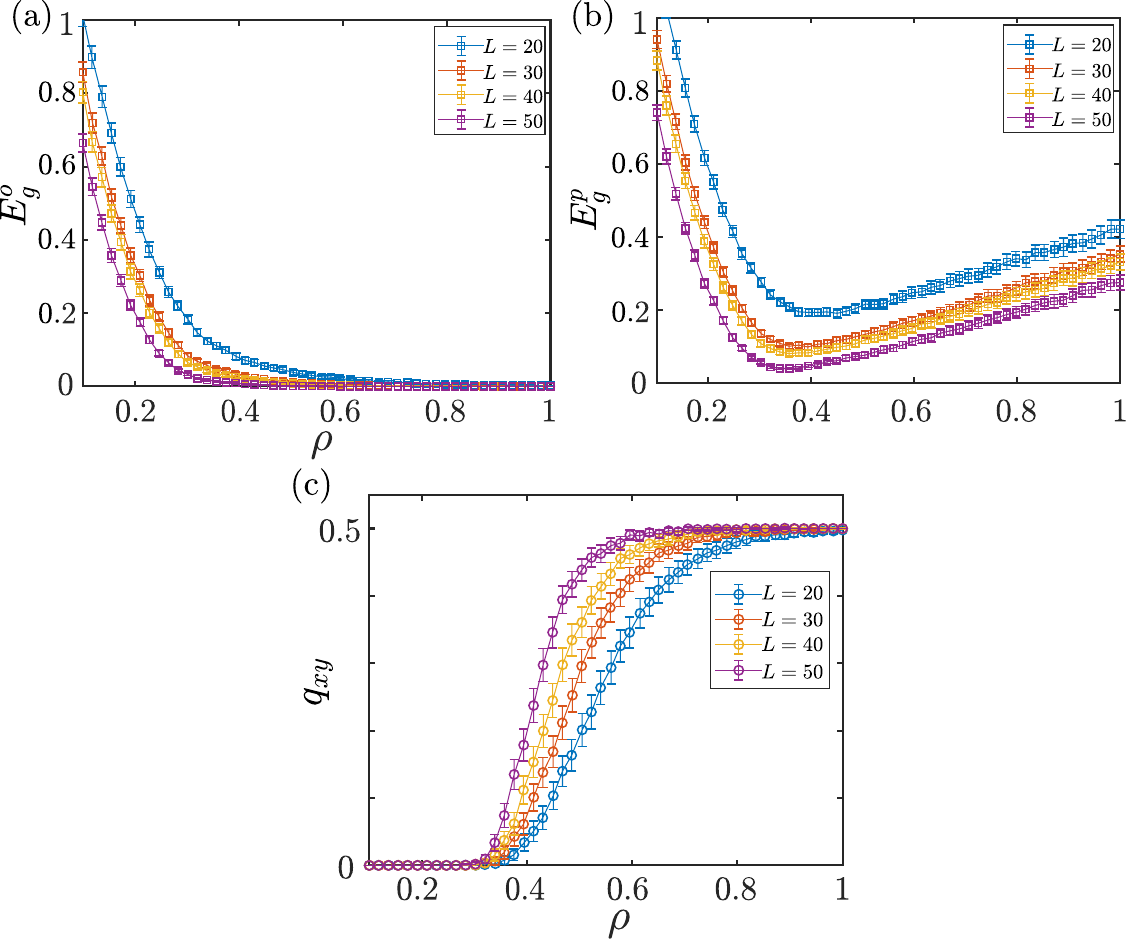} \caption{The energy gap as a function of the density with different sample sizes in (a) OBC and (b) PBC. (c) The real-space quadrupole moment $q_{xy}$ vs density with different sample sizes. We set $M=0$ and an average of $1000$ random configurations are taken.}%
\label{fig4}
\end{figure}

 In fact, the energy gap calculated in the OBC also has a rich physical meaning. To be specific, if the system is in a topological nontrivial phase, the energy gap with the OBC corresponds to the energy gap of the boundary states due to the finite-size effect. If the system is topologically trivial, the energy gap calculated in the OBC should be equal to the bulk energy gap since no boundary states exist. In Fig.~\ref{fig3}(c), the red line and orange line represent the energy gap versus density $\rho$ with OBC and PBC, respectively. It is found that the values of the energy gap with OBC and PBC are nonzero due to the system being in a topological trivial phase when the density is small ($0.1<\rho<0.28$). However, the values of the energy gap with OBC are similar to that of the energy gap with PBC but not equal with the same density, which is because the amorphous lattice is lacking periodicity and the PBC is constructed by the quasiperiodic approximation theory, inevitably bringing some additional hopping terms. We find that the values of the energy gap in both OBC and PBC decay exponentially as the density increases. Further increasing the density, the bulk energy gap has a minimum value $E_{g}^p \approx 0.096$. The corresponding critical density is $\rho \approx 0.41$, beyond which the bulk energy gap increases. This process seems to be similar to the bulk energy gap closing and then reopening, however, the minimum values of the bulk energy gap are not exactly equal to zero in our calculations. We attribute it to the finite-size effect of the system. In addition, it is found that the open boundary energy gap closes at $\rho \approx 0.55$. This is because the amorphous system is already in a higher-order topological phase and the degenerate zero-energy states appear in the energy gap. Thus, the energy gap of the boundary states is closed.

In order to get a better understanding of the finite-size effect on the topological phase transition mentioned above, we studied the energy gap $E_{g}^o$, $E_{g}^p$ and quadrupole moment $q_{xy}$ versus density $\rho$ for different sample sizes, as shown in Fig.~\ref{fig4}. It is found that the critical values of the density of the topological phase transition in an amorphous system are size dependent. More specifically, the critical density where the energy gap is closed with the OBC becomes smaller with an increase of the sample size, as shown in Fig.~\ref{fig4}(a). Meanwhile, the minimum value of the bulk energy gap becomes smaller and closer to $0$ with an increase of the sample size, as shown in Fig.~\ref{fig4}(b). It can be inferred that the minimum value of the bulk energy gap should be equal to zero in the thermodynamic limit. In addition, with an increase in the sample size, the critical density corresponding to $q_{xy}$ changing from $0$ to $0.5$ gradually decreases, as shown in Fig.~\ref{fig4}(c).

In addition, we find that the density-driven higher-order topological phase transition is also effective for a 3D system. We start from a 3D amorphous higher-order topological insulator, supporting eight corner modes, for which the Hamiltonian can be written as \cite{Benalcazar61,PhysRevB.96.245115}
\begin{eqnarray}
H_{3D} &=&-\sum_{j}c_{j}^{\dagger }(\gamma _{y}\tau _{3}\sigma
_{2}s_{2}-\gamma _{x}\tau _{1}\sigma _{0}s_{0}-\gamma _{z}\tau _{3}\sigma
_{1}s_{0})c_{j} \notag\\
&&+\sum_{j\neq k}\frac{l(r_{jk})}{2}c_{j}^{\dagger }[-i\lambda _{x}\tau _{3}\sigma
_{2}s_{3}C_{1}C_{2}+\lambda _{x}\tau _{1}\sigma _{0}s_{0}|C_{1}|C_{2} \notag\\
&&-i\lambda _{y}\tau _{3}\sigma _{2}s_{1}S_{1}C_{2}-\lambda _{y}\tau
_{3}\sigma _{2}s_{2}|S_{1}|C_{2} \notag\\
&&+\lambda _{z}\tau _{3}\sigma _{1}s_{0}|C_{1}||S_{2}|+i\lambda _{z}\tau
_{2}\sigma _{0}s_{0}|C_{1}|S_{2}]c_{k},
\label{H3D}
\end{eqnarray}
where  $c_{j}^{\dag }=(c_{j1 }^{\dag },c_{j2}^{\dag
},c_{j3 }^{\dag },c_{j4 }^{\dag },c_{j5 }^{\dag },c_{j6 }^{\dag },c_{j7 }^{\dag },c_{j8 }^{\dag })$  is the creation  operator in cell $j$. $\gamma_{x,y,z}$ and $\lambda_{x,y,z}$ are the intracell hopping amplitudes and intercell hopping amplitudes along the $x$ axis, $y$ axis, and $z$ axis, respectively. $S_{1}=\sin (\phi _{jk})$, $S_{2}=\sin (\theta _{jk})$, $C_{1}=\cos (\phi _{jk})$, and $C_{2}=\cos (\theta _{jk})$, where $\phi _{jk}$ and $\theta _{jk}$ are the azimuth and elevation angles, respectively. $\tau$, $\sigma$, and $s$ are Pauli matrices for the degrees of freedom within a unit cell. We set $\gamma_{x,y,z}=-0.25$ and  $\lambda_{x,y,z}=1$ for numerical simulation.

\begin{figure}[htp]
	\includegraphics[width=8.5cm]{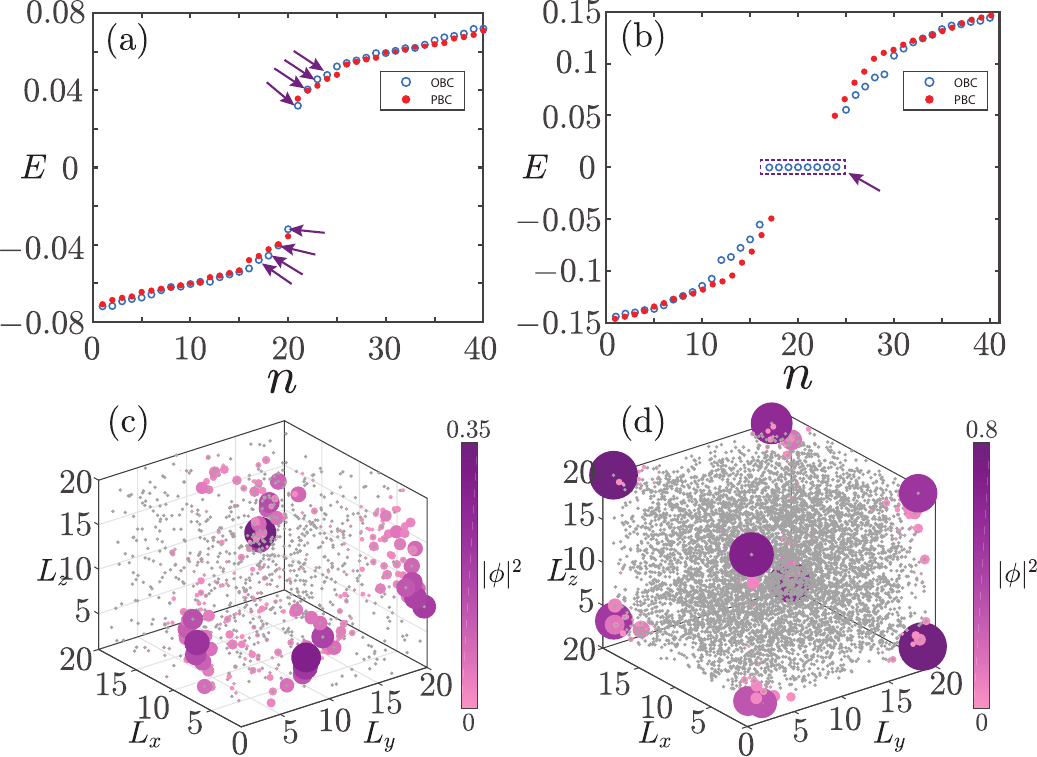} \caption{Energy spectrum and the probability density of the eight eigenstates which are the nearest to zero energy (marked by purple arrows) at OBC with different densities. (a), (c) $\rho=0.125$ and (b), (d) $\rho=1$.  The sample size is $L_{x}\times L_{y}\times L_{z}=20\times20\times20$. We set the cutoff distance $R=2.5$.}%
\label{fig5}
\end{figure}

In Figs.~\ref{fig5}(a) and \ref{fig5}(b), we plot the energy spectrum of the $3$D amorphous lattice with different densities $\rho$ in both OBC (marked by blue circles) and PBC (marked by red dots). For the case of $\rho=0.125$, the system is a topological trivial phase with a small energy gap. The corresponding wave-function distributions of the eight eigenstates which are the nearest to zero energy [marked by purple arrows in Fig.~\ref{fig5}(a)] are localized in the bulk, as shown in Fig.~\ref{fig5}(c). For another case of $\rho=1$, eight zero-energy modes appear in the energy gap. The corresponding wave-function distributions of the eight zero-energy modes [marked by the purple arrow in Fig.~\ref{fig5}(b)] are localized at the eight corners of the lattice, indicating the system is in a higher-order topological phase. Thus, it is confirmed that a density-driven topological phase transition from a topological trivial phase to a higher-order topological phase can be realized in $3$D amorphous systems. Actually, the real-space octupolar moment is a well-defined topological invariant which can characterize the higher-order topological phase in 3D amorphous systems. However, due to our limited computing power at this stage, we cannot satisfy the need of computing  the topological invariants of 3D systems and will continue to study it in the future.

\section{Conclusion and discussion}
\label{CD}
In this paper, we investigate the density-driven higher-order topological phase transition in a $2$D amorphous lattice. Based on calculating the quadrupole moment and detecting the presence of corner states, we demonstrate that a higher-order topological insulator phase can exist in an entire random lattice. More interestingly, the topological phase transition from a normal phase to a higher-order topological insulator phase occurs in the amorphous system with increasing density. In addition, the density can also drive a higher-order topological phase transition in a $3$D amorphous lattice.

It is well known that higher-order topological insulators are established as topological crystalline insulators protected by crystalline symmetries. However, with further study, it is found that higher-order topological insulators can also exist in aperiodic systems, such as quasicrystals and amorphous solids, which lack translational symmetry. In our model, higher-order topological insulators are protected by particle-hole symmetry and effective chiral symmetry. In fact, the geometric structure of a $2$D ($3$D) amorphous lattice can be constructed by adding structural disorder to a square (cubic) lattice, so that each site in a square (cubic) lattice generates a random deviation. Thus, an amorphous system can be regarded as a clean system with structural disorder in some sense. Very recently, structural disorder-induced first-order and second-order topological phase transitions have been proposed \cite{PhysRevLett.128.056401,PhysRevLett.126.206404}. According to our calculations, it is found that density can drive higher-order topological phase transitions in amorphous systems. We suppose that an increase in density brings some additional hopping terms into the system, leading to the occurrence of a topological phase transition. In addition, changing the density of the amorphous system is equivalent to modulating the strength of the structural disorder, thus, density-driven topological phase transitions are reasonable in amorphous systems. Moreover, we find that our phase diagram in the 2D model correlates very well with Ref. \cite{costa2021discovery}. In some sense, the density in our model has a similar physical effect as the hopping parameters for the higher-order topological phase transitions.

Recently, $3$D amorphous topological insulators protected by time-reversal symmetry have been experimentally observed in Bi$_{2}$Se$_{3}$ films \cite{corbae2021evidence}. In addition, a density-driven phase transition between semiconducting and metallic polyamorphs of silicon has been realized by changing pressure \cite{mcmillan2005density}. Therefore, we propose that density-driven higher-order topological phase transitions may be realized in amorphous Bi$_{2}$Se$_{3}$ thin films by changing the pressure of the sample.

\section*{Acknowledgments}
B.Z. was supported by the NSFC (under Grant No. 12074107) and the program of outstanding young and middle-aged scientific and technological innovation team of colleges and universities in Hubei Province (under Grant No. T2020001). H.-M.H. was supported by Science and Technology Innovation Team in Colleges of Hubei Province (under Grant No. T2021012).

\bibliographystyle{apsrev4-1-etal-title_6authors}
\end{document}